\documentclass[12pt,cite,epsf,epsfig]{article}
\usepackage{epsfig}
\usepackage[dvips]{color}

\begin{document}

\title{Leptogenesis in a Hybrid Texture Neutrino Mass Model}
\author{{S. Dev}{\thanks{{Electronic address: dev5703@yahoo.com}} and}
{Surender Verma}{\thanks{{Electronic address: s\_7verma@yahoo.co.in}}}}
\date{\textit{{Department of Physics, Himachal Pradesh University, Shimla 171005, INDIA.}}}
\maketitle

\begin{abstract}
We investigate the CP asymmetry for a hybrid texture of the neutrino mass matrix predicted by $Q_8$ family symmetry in the context of the type-I seesaw mechanism and examine its consequences for leptogenesis. We, also, calculate the resulting Baryon Asymmetry of the Universe (BAU) for this texture.
\end{abstract}

\section{INTRODUCTION}
In the Standard Model (SM), fermions acquire masses via
spontaneous breakdown of SU(2) gauge symmetry. However, the values
of fermion masses and the observed hierarchical fermion spectra
are not understood within the SM. This results in thirteen free
parameters in the SM which includes three charged lepton masses,
six quark masses and the four parameters of the CKM matrix. The
symmetries of the SM do not allow non-zero neutrino masses through
renormalizable Yukawa couplings. However, non-zero neutrino masses
can be introduced via non-renormalizable higher dimensional
operators presumably having their origin in physics beyond the SM.
Radiative and seesaw mechanisms often supplimented by additional inputs like texture zeros and flavor symmetries are widely discussed mechanisms for fermion mass generation. These mechanisms, most often, complement
and reinforce each other. In the ongoing decade, significant
advances have been made in understanding these mechanisms. In
particular texture zeros and flavor symmetries have provided
quantitative relationships between flavor mixing angles and the
quark/lepton mass ratios. It has, now been realized that the
``See-Saw GUT'' scenario, on its own, cannot provide a complete
understanding of the flavor structure of the quark and lepton mass
matrices and new physics seems to be essential perhaps in the form
of new symmetries mainly in the lepton sector. Moreover, a unified
description of flavor physics and CP-violation in the quark and
lepton sectors is absolutely necessary. This can be achieved by
constructing a low energy effective theory with the SM and some
discrete non-Abelian family symmetry and, subsequently, embedding
this theory into Grand Unified Theory (GUT) models like
SO(10)\cite{1}. For this reason, the discrete symmetry will have
to be a subgroup of SO(3) or SU(3)\cite{2}. The search for an
adequate discrete symmetry has mainly focussed on the minimal
subgroups of these groups with at least one singlet and one
doublet irreducible representation to accommodate the fermions
belonging to each generation. One such subgroup is the quaternion
group $Q_8$\cite{3} which not only accommodates the three
generations of fermions but also explains the rather large
difference between values of 2-3 mixings in the quark and lepton
sectors.

Discrete quaternion groups have been extensively studied in the literature\cite{3}. $Q_8$ is a subgroup of $SU(2)$.
A two-dimensional representation of $Q_8$ is provided by the eight unimodular
unitary matrices ($\pm \textbf{1}$, $\pm i\sigma_{1}$, $\pm i\sigma_{2}$, $\pm i\sigma_{3}$) 
where $\textbf{1}$ is the $2\times2$ unit matrix and ($\sigma_{1}$, $\sigma_{2}$, $\sigma_{3}$) are the Pauli Matrices. There are five irreducible representations of $Q_8$ corresponding to the five conjugacy classes 
in which the elements of $Q_8$ are divided and are conveniently denoted by $\textbf{1}^{++}$, $\textbf{1}^{+-}$, $\textbf{1}^{-+}$, $\textbf{1}^{--}$ and $\textbf{2}$. The SU(2) decomposition of these irreducible representations is given by $\textbf{1}_{SU(2)}=\textbf{1}^{++}$, $\textbf{2}_{SU(2)}=\textbf{2}$, $\textbf{3}_{SU(2)}=\textbf{1}^{+-} + \textbf{1}^{-+} + \textbf{1}^{--}$. The irreducible representations $\textbf{1}^{+-}$, $\textbf{1}^{-+}$ and $\textbf{1}^{--}$ having the same group properties are equivalent and are, hence, interchangeable. The basic tensor product rule for $Q_8$ is $2\times2$ =$\underbrace{\textbf{1}^{++}}_{singlet}+\underbrace{\textbf{1}^{+-}+\textbf{1}^{-+}+\textbf{1}^{--}}_{triplet} $. The quarks and leptons can be differentiated by assigning them to different irreducible representations of $Q_8$ in the following manner:

\begin{equation}
(u_{\alpha}, d_{\alpha}), u_{\alpha}^{c}, d_{\alpha}^{c} \in \textbf{1}^{--}, \textbf{1}^{-+}, \textbf{1}^{+-},
\end{equation}
\begin{equation}
(\nu_{\alpha}, l_{\alpha}), l_{\alpha}^{c} \in \textbf{1}^{++}, \textbf{2}
\end{equation}
where $\alpha$ is the flavor index.

The Higgs doublets have the following $Q_8$ assignments:
\begin{equation}
(\phi_{1}^{o}, \phi_{1}^{-}) \sim \textbf{1}^{++},
\end{equation}

\begin{equation}
(\phi_{2}^{o}, \phi_{2}^{-}) \sim \textbf{1}^{+-}.
\end{equation}
Electroweak symmetry breaking generates the charged fermion mass terms $M_{f}^{ij}f_{i}f_{j}$ and the neutrino mass terms $M_{\nu}^{ij}\nu_{i}\nu_{j}$. The entries in the fermion mass matrices are associated with $Q_8$ 
assignments of the corresponding fermion bilinears leading to

 \begin{equation}
 M_{quark} \sim \left(%
\begin{array}{ccc}
  \textbf{1}^{++} & \textbf{1}^{+-} & \textbf{1}^{-+} \\
  \textbf{1}^{+-} & \textbf{1}^{++} & \textbf{1}^{--} \\
  \textbf{1}^{-+} & \textbf{1}^{--} & \textbf{1}^{++} \\
\end{array}%
\right)
\end{equation}
and 

\begin{equation}
 M_{lepton} \sim \left(%
\begin{array}{ccc}
  \textbf{1}^{++} & \textbf{2} & \textbf{2} \\
  \textbf{2} & \textbf{1}^{--}+\textbf{1}^{+-} & \textbf{1}^{-+}+\textbf{1}^{++} \\
  \textbf{2} & \textbf{1}^{-+}-\textbf{1}^{++} & \textbf{1}^{--}-\textbf{1}^{+-} \\
\end{array}%
\right).
\end{equation}

The $\textbf{1}^{++}$ contribution in the 2-3 sector of the Majorana neutrino mass matrix is forbidden because of the symmetry of the mass matrix. Nonzero entries in the neutrino mass matrix are induced by the vacuum expectation values (VEVs) of Higgs triplets transforming in the corresponding $Q_8$ irreducible representation. These triplet VEVs are induced by trilinear couplings of the type $\xi_{i}\varphi_{j}\varphi_{k}$\cite{4}. Two Higgs doublets generate the masses for charged fermions. At least four Higgs triplets are needed to correctly reproduce the current neutrino data. Nonzero mixing in the solar sector requires the VEVs of $(\xi_{3}, \xi_{4})\in \textbf{2}$. The other two Higgs triplets $\xi_{1}$ and $\xi_{2}$ must transform in two different 1-dimensional representations of $Q_{8}$ for which four distinct choices are possible corresponding to $Q_{8}$ assignments relative to $(\phi_{1}^{o}, \phi_{1}^{-}) \in \textbf{1}^{++}$ and $(\phi_{2}^{o}, \phi_{2}^{-}) \in \textbf{1}^{+-}$ viz. (i) $\xi_{1} \in \textbf{1}^{++}, \xi_{2} \in \textbf{1}^{+-}$ (ii) $\xi_{1} \in \textbf{1}^{++}, \xi_{2} \in \textbf{1}^{-+}/\textbf{1}^{--}$ (iii) $\xi_{1} \in \textbf{1}^{-+}/\textbf{1}^{--}, \xi_{2} \in \textbf{1}^{+-}$ (iv) $\xi_{1} \in \textbf{1}^{-+}, \xi_{2} \in \textbf{1}^{--}$.

The resulting Majorana neutrino mass matrices\cite{3} corresponding to the above $Q_8$ assignments are

 \begin{equation}
 M_{\nu}^{I}=\left(%
\begin{array}{ccc}
  a & c & d \\
  c & 0 & b \\
  d & b & 0 \\
\end{array}%
\right),
 M_{\nu}^{II}=\left(%
\begin{array}{ccc}
  0 & c & d \\
  c & a & 0 \\
  d & 0 & b \\
\end{array}%
\right),
\end{equation}

\begin{equation}
M_{\nu}^{III}=\left(%
\begin{array}{ccc}
  0 & c & d \\
  c & a & b \\
  d & b & a \\
\end{array}%
\right),
M_{\nu}^{IV}=\left(%
\begin{array}{ccc}
  a & c & d \\
  c & b & 0 \\
  d & 0 & b \\
\end{array}%
\right) 
 \end{equation}
 in the charged lepton basis.
  Out of these four possible scenarios for the Majorana
  neutrino mass matrix, the scenario $II$ and $IV$ are excluded by the
  present experimental data on neutrino masses and mixings\cite{3,5}.
  In fact, scenario $I$ and $II$ correspond to two texture zero neutrino mass
  matrices studied extensively in the literarture\cite{6,7,8,9,10,11,12,13,14} and leptogenesis in these two scenarios have been investigated\cite{15,16}. However, the hybrid scenario
  $III$ is a new scenario predicted by the $Q_8$
   symmetry with a normal hierarchy of neutrino masses and non-zero 1-3
   mixing angle\cite{5,17}. In this scenario, equalities between mass matrix elements can coexist with texture zeros. It is extremely important to investigate the connection between the textures of fermion mass matrices and the observables of flavor mixing. Discrete quaternion groups have been extensively applied to flavor physics\cite{18,19,20,21,22}. It is, therefore, important to subject models based on $Q_8$, in particular, to the test of a viable leptogenesis to explain the Baryon Asymmetry of the Universe (BAU). Leptogenesis is based on the $CP$ asymmetry generated through out of equilibrium lepton number violating decays of heavy Majorana neutrinos resulting in a lepton asymmetry which is subsequently transformed into a baryon asymmetry via $(B+L)$ violating sphaleron processes and depends on the structure of Majorana neutrino mass matrix. In the present work, we study the CP asymmetry in the hybrid scenario $III$ predicted by $Q_8$ symmetry and examine its implications for leptogenesis and the resulting Baryon Asymmetry of the Universe (BAU).
   \section{Neutrino Mass Matrix}
   We consider the following neutrino mass matrix with a hybrid
   texture resulting from $Q_8$ symmetry:
 \begin{equation}
 M_{\nu}=\left(%
\begin{array}{ccc}
  0 & a & b \\
  a & c & d \\
  b & d & c \\
\end{array}%
\right)
 \end{equation}
 where $a$, $b$, $c$ and $d$ are complex, in general. In the framework of type-I seesaw mechanism\cite{23,24,25,26},
 the effective light neutrino mass matrix $M_{\nu}$ is given by
\begin{equation}
 M_{\nu}=M_DM_R^{-1}M_D^T
 \end{equation}
 where $M_D$ is the Dirac neutrino mass matrix and $M_R$ is the
 right-handed Majorana neutrino mass matrix. The number of physical seesaw parameters contained in $M_{D}$ and $M_{R}$ on the right hand side is double the number of parameters in the low energy neutrino mass matrix $M_{\nu}$. As a result, the reconstruction of the seesaw is not possible solely from low energy neutrino physics and one requires additional observables. One such observable which can be used in reconstructing the seesaw is the observed Baryon Asymmetry of the Universe (BAU). The seesaw matrices $M_{D}$ and $M_{R}$ are totally unknown and even the light neutrino mass matrix $M_{\nu}$ is not completely known. In the absence of a complete knowledge of $M_{\nu}$ and $M_{R}$, there are an infinite number of possibilities for the seesaw Dirac neutrino mass matrix $M_{D}$ resulting in the so-called seesaw degeneracy\cite{27}. Therefore, to facilitate the reconstruction of the seesaw, one is constrained to make additional assumptions about the seesaw matrices motivated by some specific models. However, in general, in the absence of a specific model in hand, the best one can do is to parametrize one's ignorance and that was the approach followed by Casas and Ibarra culminating in the so-called Casas-Ibarra (CI) parametrization\cite{28} given by
 \begin{equation}
 M_D=iV\sqrt{M_{\nu}^d}R\sqrt{M_{R}^d}
 \end{equation}
where $M_{\nu}^d=diag \lbrace{m_1,m_2,m_3}\rbrace$ ($m_i, i=1,2,3$ are light neutrino masses), $M_{R}^d=diag \lbrace{M_1,M_2,M_3}\rbrace$ ($M_i, i=1,2,3$ are right handed Majorana neutrino masses), $V$ neutrino mixing matrix and $R$ is a complex orthogonal matrix.  The light neutrino mass matrix $M_{\nu}$ (from low
energy phenomenology) can be parametrized as
\begin{equation}
M_{\nu}=m_0\left (\begin{array}{ccc}
  0 & \lambda & \lambda \\
  \lambda & 1 & \epsilon-1 \\
  \lambda & \epsilon-1 & 1 \\
\end{array}\right)
\end{equation}
where the real parameters $\lambda$, $\epsilon$ and $m_0$ are
given by $\lambda=0.01$, $\epsilon=0.12$ and
$m_0$=$\frac{\sqrt{\Delta m^2_{23}}}{2}$ with $\Delta m_{23}^{2} =
2.37\times 10^{-3}$ eV$^{2}$\cite{29} and the apparent equalities between different elements of the neutrino mass matrix are to within the current precision ($\approx 10^{-2}$) of the oscillation parameters. One can diagonalize $M_{\nu}$ given by Eqn. (12) to calculate matrices $M_{\nu}^d$ and $V$:
\begin{equation} 
M_{\nu}^d=\left (\begin{array}{ccc}
  m_1 & 0 & 0 \\
  0 & m_2 & 0 \\
  0 & 0 & m_3 \\ 
\end{array}\right)
\end{equation}
\begin{equation}
=m_0\left (\begin{array}{ccc}
  \frac{1}{2} \left(\epsilon -\sqrt{\epsilon ^2+8 \lambda ^2}\right) & 0 & 0 \\
  0 & \frac{1}{2} \left(\sqrt{\epsilon ^2+8 \lambda ^2}+\epsilon \right)& 0 \\
  0 & 0 & 2-\epsilon \\
\end{array}\right),
\end{equation}
and
\begin{equation}
V=\left(
\begin{array}{ccc}
 -\frac{\epsilon +\sqrt{\epsilon ^2+8 \lambda ^2}}{\sqrt{8 \lambda ^2+\left(\epsilon +\sqrt{\epsilon ^2+8 \lambda ^2}\right)^2}} &
   \frac{\sqrt{\epsilon ^2+8 \lambda ^2}-\epsilon }{\sqrt{8 \lambda ^2+\left(\epsilon -\sqrt{\epsilon ^2+8 \lambda ^2}\right)^2}} &
   0 \\
 \frac{2 \lambda }{\sqrt{8 \lambda ^2+\left(\epsilon +\sqrt{\epsilon ^2+8 \lambda ^2}\right)^2}} & \frac{2 \lambda }{\sqrt{8
   \lambda ^2+\left(\epsilon -\sqrt{\epsilon ^2+8 \lambda ^2}\right)^2}} & -\frac{1}{\sqrt{2}} \\
 \frac{2 \lambda }{\sqrt{8 \lambda ^2+\left(\epsilon +\sqrt{\epsilon ^2+8 \lambda ^2}\right)^2}} & \frac{2 \lambda }{\sqrt{8
   \lambda ^2+\left(\epsilon -\sqrt{\epsilon ^2+8 \lambda ^2}\right)^2}} & \frac{1}{\sqrt{2}}
\end{array}
\right).
\end{equation}
 The complex orthogonal matrix $R$ can be parameterized as
\begin{equation}
R=\left(\begin{array}{ccc}
   R_{11} & R_{12} & R_{13} \\
   R_{21} & R_{22} & R_{23} \\
   R_{31} & R_{32} & R_{33} \\
    \end{array}
    \right)=T_{12}T_{13}T_{23},
\end{equation}
where $T_{ij}$ is the matrix of rotation by a complex angle $\xi_{ij}=\eta_{ij}+i\zeta_{ij}$ in the $ij$- plane with

\begin{equation}
 T_{12}=\left(\begin{array}{ccc}
   c_{12} & s_{12} & 0 \\
   -s_{12} & c_{12} & 0 \\
   0 & 0 & 1 \\
    \end{array}
    \right); T_{23}=\left(\begin{array}{ccc}
   1 & 0 & 0 \\
   0 & c_{23} & s_{23} \\
   0 & -s_{23} & c_{23} \\
    \end{array}
    \right); T_{13}=\left(\begin{array}{ccc}
   c_{13} & 0 & s_{13} \\
   0 & 1 & 0 \\
   -s_{13} & 0 & c_{13} \\
    \end{array}
    \right),
 \end{equation}
 where $c_{ij}=\cos \xi_{ij}$, $s_{ij}=\sin \xi_{ij}$ for $i<j$.
 Using Eqns. (14-17) the Dirac neutrino mass matrix $M_D$ in the CI parameterization can be written as:
\begin{equation}
M_{D}=\left (\begin{array}{ccc}
  M_{D}^{11} & M_{D}^{12} & M_{D}^{13} \\
  M_{D}^{21} & M_{D}^{22} & M_{D}^{23} \\
  M_{D}^{31} & M_{D}^{32} & M_{D}^{33} \\
\end{array}\right),
\end{equation}
where elements of $M_{D}$ are given by
\begin{equation}
 \left.\begin{array}{l}

M_{D}^{11}=i \sqrt{M_1} \left(-A_2 c_{13} \sqrt{m_2} s_{12}-A_1 c_{12} c_{13} \sqrt{m_1}\right)\\
M_{D}^{12}=i \sqrt{M_2} \left(A_2 \sqrt{m_2} \left(c_{12} c_{23}+s_{12} s_{13} s_{23}\right)-A_1 \sqrt{m_1} \left(c_{23} s_{12}-c_{12} s_{13}
   s_{23}\right)\right)\\
M_{D}^{13}=i \sqrt{M_3} \left(A_2 \sqrt{m_2} \left(c_{12} s_{23}-c_{23} s_{12} s_{13}\right)-A_1 \sqrt{m_1} \left(c_{12} c_{23} s_{13}+s_{12}
   s_{23}\right)\right)\\
M_{D}^{21}=i \sqrt{M_1} \left(-\frac{2 A_2 c_{13} \sqrt{m_2} s_{12} \lambda }{\sqrt{\epsilon ^2+8 \lambda ^2}-\epsilon }+\frac{2 A_1 c_{12}
   c_{13} \sqrt{m_1} \lambda }{\sqrt{\epsilon ^2+8 \lambda ^2}+\epsilon }+\frac{\sqrt{m_3} s_{13}}{\sqrt{2}}\right)\\
M_{D}^{22}=i \sqrt{M_2} \left(\frac{2 A_1 \sqrt{m_1} \lambda  \left(c_{23} s_{12}-c_{12} s_{13} s_{23}\right)}{\sqrt{\epsilon ^2+8 \lambda
   ^2}+\epsilon }+\frac{2 A_2 \sqrt{m_2} \lambda  \left(c_{12} c_{23}+s_{12} s_{13} s_{23}\right)}{\sqrt{\epsilon ^2+8 \lambda
   ^2}-\epsilon }+\frac{c_{13} \sqrt{m_3} s_{23}}{\sqrt{2}}\right)\\
M_{D}^{23}=i \sqrt{M_3} \left(\frac{2 A_2 \sqrt{m_2} \lambda  \left(c_{12} s_{23}-c_{23} s_{12} s_{13}\right)}{\sqrt{\epsilon ^2+8 \lambda
   ^2}-\epsilon }+\frac{2 A_1 \sqrt{m_1} \lambda  \left(c_{12} c_{23} s_{13}+s_{12} s_{23}\right)}{\sqrt{\epsilon ^2+8 \lambda
   ^2}+\epsilon }-\frac{c_{13} c_{23} \sqrt{m_3}}{\sqrt{2}}\right)\\
M_{D}^{31}=i \sqrt{M_1} \left(-\frac{2 A_2 c_{13} \sqrt{m_2} s_{12} \lambda }{\sqrt{\epsilon ^2+8 \lambda ^2}-\epsilon }+\frac{2 A_1 c_{12}
   c_{13} \sqrt{m_1} \lambda }{\sqrt{\epsilon ^2+8 \lambda ^2}+\epsilon }-\frac{\sqrt{m_3} s_{13}}{\sqrt{2}}\right)\\
M_{D}^{32}=i \sqrt{M_2} \left(\frac{2 A_1 \sqrt{m_1} \lambda  \left(c_{23} s_{12}-c_{12} s_{13} s_{23}\right)}{\sqrt{\epsilon ^2+8 \lambda
   ^2}+\epsilon }+\frac{2 A_2 \sqrt{m_2} \lambda  \left(c_{12} c_{23}+s_{12} s_{13} s_{23}\right)}{\sqrt{\epsilon ^2+8 \lambda
   ^2}-\epsilon }-\frac{c_{13} \sqrt{m_3} s_{23}}{\sqrt{2}}\right)\\
M_{D}^{33}=i \sqrt{M_3} \left(\frac{2 A_2 \sqrt{m_2} \lambda  \left(c_{12} s_{23}-c_{23} s_{12} s_{13}\right)}{\sqrt{\epsilon ^2+8 \lambda
   ^2}-\epsilon }+\frac{2 A_1 \sqrt{m_1} \lambda  \left(c_{12} c_{23} s_{13}+s_{12} s_{23}\right)}{\sqrt{\epsilon ^2+8 \lambda
   ^2}+\epsilon }+\frac{c_{13} c_{23} \sqrt{m_3}}{\sqrt{2}}\right)\\

\end{array}  \right\}
\end{equation}
and the quantities $A_1$ and $A_2$ are given by
\begin{eqnarray}
A_1=\frac{\sqrt{\epsilon ^2+8 \lambda ^2}+\epsilon }{\sqrt{\left(\sqrt{\epsilon ^2+8 \lambda ^2}+\epsilon \right)^2+8 \lambda ^2}}, \nonumber \\
   A_2=\frac{\sqrt{\epsilon ^2+8 \lambda ^2}-\epsilon }{\sqrt{\left(\epsilon -\sqrt{\epsilon ^2+8 \lambda ^2}\right)^2+8 \lambda ^2}}.
\end{eqnarray}
\section{Hybrid Texture and Baryon Asymmetry}
Baryon Asymmetry of the Universe (BAU) poses a puzzle for particle
physics as well as cosmology. Even though the Standard Model (SM)
has all the ingredients\cite{30,31,32,33} necessary for the
dynamical generation of baryon asymmetry, it fails to explain the
observed baryon asymmetry as the SM CP-violation is too small to
generate the observed baryon asymmetry. Moreover, the electroweak
phase transition (EWPT) is not strongly first order as required
for successful baryogenesis. Baryogenesis, thus, requires new
physics beyond the SM essentially in the form of new sources of
CP-violation and must either provide a departure from thermal
equilibrium in addition to the electroweak phase transition or
modify the electroweak phase transition itself. Some of the
possible new physics mechanisms are Affleck-Dine
mechanism\cite{34,35}, GUT baryogenesis\cite{36,37,38,39,40,41,42,43,44,45} and
baryogenesis via leptogenesis\cite{46,47} etc.. Out of these
scenarios, the last one is particularly appealing since there is a
plethora of reasons to believe that the SM is only a low energy
effective theory and there are strong indications of new physics
at a higher energy scale. The experimental evidence for massive
neutrinos, the dark matter puzzle apart from the fine tuning
problem of the Higgs mass and the gauge coupling unification are
some of these reasons for invoking physics beyond the SM. It is,
particularly interesting to note that the mechanism of
baryogenesis via leptogenesis is motivated by some of the reasons
listed above. In this mechanism, heavy singlet neutrinos are
introduced via the seesaw mechanism whose Yukawa couplings provide
new sources of CP-violation essential for a viable leptogenesis.
The rates of these new Yukawa interactions can be slow enough to
generate departure from thermal equilibrium. Majorana masses of
heavy singlet neutrinos lead to necessary lepton number violation
and the (B+L) violating SM sphaleron processes\cite{30} play a
crucial role in partially converting the lepton asymmetry into a
net baryon asymmetry. In this section we calculate the baryon
asymmetry in the hybrid texture model (Eqn. (9)). The physical
Majorana neutrino, $N_R$, decays in two modes:
\begin{eqnarray}
N_R\rightarrow l_{\alpha}+\phi^\dagger \\
\rightarrow \overline{l_{\alpha}}+\phi
\end{eqnarray}
where $l_{\alpha}$ is lepton and $\overline{l_{\alpha}}$ is antilepton. The CP
asymmetry which is caused by the interference of tree level with
one loop corrections for the decays of the lightest of heavy right
handed Majorana neutrino $N_1$ is given by\cite{48,49}

\begin{eqnarray}
\varepsilon_{1}^{\alpha}=\frac{\Gamma-\overline{\Gamma}}{\Gamma+\overline{\Gamma}}
=\frac{1}{8\pi
v^2}\frac{1}{\left( M_D^\dagger{M_D}\right)_{ii}}
\sum_{j\neq
i} \Im\left[\left(M^{\dagger}_{D}\right)_{i\alpha}\left(M_{D}\right)_{\alpha j} \left(M^{\dagger}_{D}M_{D} \right)_{ij}  \right]  f\left(\frac{M_j^2}{M_i^2}\right)+\nonumber \\
\frac{1}{8\pi
v^2}\frac{1}{\left( M_D^\dagger{M_D}\right) _{ii}}
\sum_{j\neq
i} \Im\left[\left(M^{\dagger}_{D}\right)_{i\alpha}\left(M_{D}\right)_{\alpha j} \left(M^{\dagger}_{D}M_{D} \right)_{ji}  \right]\frac{1}{1-\frac{M_{j}^2}{M_{i}^2}}
\end{eqnarray}

where $\Gamma=\Gamma(N_1\rightarrow l_{\alpha}\phi^\dagger)$ and
$\overline{\Gamma}=\Gamma(N_1\rightarrow \overline{l_{\alpha}}\phi)$ are
the decay rates and $v$ is the scale of the electroweak symmetry
breaking, $v\simeq174$ GeV. Within the SM the function $f(\varsigma)$ has the form
\begin{equation}
f(\varsigma)=\sqrt{\varsigma}\left[ \frac{1}{1-\varsigma}+1-(1+\varsigma)ln(\frac{1+\varsigma}{\varsigma})\right],
\end{equation}
where $\varsigma=\frac{M_{j}^2}{M_{i}^2}$.

 For hierarchical right handed Majorana neutrino masses
i.e. $M_1 \ll M_2$, $M_3$ the function $f(\varsigma)\simeq\frac{-3}{2\sqrt{\varsigma}}$ and, in addition, the second term in Eqn. (23) is strongly suppressed, therefore, we will neglect this term in the following analysis. Also, the CP asymmetry given by Eqn. (23) when summed over the flavors $\alpha$ ($M_1 \geq 10^{12}$ GeV) can be written as
\begin{equation}
\varepsilon_1=\sum_\alpha \varepsilon_{1}^{\alpha} \approx -\frac{3}{16\pi
v^2}\left(\frac{\Im\left[ \left( {M_D}^\dagger{M_D}\right) _{12}^2\right] }{\left( {M_D}^\dagger{M_D}\right) _{11}}
\frac{M_1}{M_2}+\frac{\Im\left[ \left( {M_D}^\dagger{M_D}\right) _{13}^2\right] }{\left( {M_D}^\dagger{M_D}\right) _{11}}
\frac{M_1}{M_3}\right).
\end{equation}
It is evident from Eqn. (25) that for hierarchical right handed Majorana neutrinos, a non-vanishing decay asymmetry $\varepsilon_1$ depends on the imaginary part of $(1,2)$ and $(1,3)$ elements of $M^{\dagger}_{D}M_{D}$ which in the Casas-Ibarra (CI) parameterization is given by
\begin{equation}
M^{\dagger}_{D}M_{D}=\sqrt{M^d_{R}}R^{\dagger}M^d_{\nu}R\sqrt{M^d_{R}}.
\end{equation}

The elements $\left( {M_D}^\dagger{M_D}\right) _{11}$,
$\left( {M_D}^\dagger{M_D}\right) _{12}$ and
$\left( {M_D}^\dagger{M_D}\right) _{13}$ are given by

\begin{eqnarray}
\left( M_D^\dagger M_D\right) _{11}=\frac{1}{2} m_0 M_1 (R_{11} (R_{11}){}^* (\epsilon -\sqrt{\epsilon ^2+8 \lambda ^2})+R_{21}
   (R_{21}){}^* (\sqrt{\epsilon ^2+8 \lambda ^2}+\epsilon )\nonumber \\   -2 R_{31} (R_{31}){}^* (\epsilon
   -2)),
\end{eqnarray}

\begin{eqnarray}
\nonumber \left( M_D^\dagger M_D\right) _{12}=\frac{1}{2} m_0 \sqrt{M_1} \sqrt{M_2} (R_{12} (R_{11}){}^* (\epsilon -\sqrt{\epsilon ^2+8 \lambda
   ^2})+R_{22} (R_{21}){}^* (\sqrt{\epsilon ^2+8 \lambda ^2}+\epsilon ) \nonumber \\ -2 R_{32}
   (R_{31}){}^* (\epsilon -2)),
\end{eqnarray}

\begin{eqnarray}
\nonumber \left( M_D^\dagger M_D\right) _{13}=\frac{1}{2} m_0 \sqrt{M_1} \sqrt{M_3} (R_{13} (R_{11}){}^* (\epsilon -\sqrt{\epsilon ^2+8 \lambda
   ^2})+R_{23} (R_{21}){}^* (\sqrt{\epsilon ^2+8 \lambda ^2}+\epsilon ) \nonumber \\ -2 R_{33}
   (R_{31}){}^* (\epsilon -2)).
\end{eqnarray}

The lepton asymmetry is related to the CP asymmetry through the
relation
\begin{equation}
Y_L=\frac{n_L-\overline{n_L}}{s}=\kappa\frac{\varepsilon_1}{g_*}
\end{equation}
where $n_L$ and $\overline{n_L}$ are number densities of leptons
and antileptons, respectively, $s$ is the entropy density,
$\kappa$ is the dilution factor which accounts for the washout
processes such as inverse decay and lepton number violating
scattering and $g_*$ is the effective number of degrees of
freedom, $g_*=106.75$\cite{15}. The possibility of generating an asymmetry between number
of leptons and antileptons (lepton asymmetry) is due to a
non-vanishing CP asymmetry, $\varepsilon_1$. The lepton asymmetry,
thus, produced is converted into a net baryon asymmetry, $Y_B$,
through the sphaleron processes which is given by the
relation\cite{50,51}
\begin{equation}
Y_B=\frac{\zeta}{\zeta-1}Y_L,
\zeta=\frac{8N_f+4N_H}{22N_f+13N_H}
\end{equation}
where $N_f$ is number of fermion families and $N_H$ is number of
complex Higgs doublets. Taking $N_f=3$ and $N_H=1$, we get
\begin{equation}
Y_B\simeq-\frac{28}{51} Y_L.
\end{equation}
In order to calculate the baryon asymmetry we need the dilution
factor $\kappa$ which involves integration over the full set of
Boltzmann equations\cite{52,53}. The approximate value of
the dilution factor which is sufficient for our purpose is given by\cite{54,55,56}
\begin{equation}
\kappa\simeq\frac{2 \left(1-e^{-\frac{1}{2} K_R Z_B(K_R)}\right)}{K_R Z_B(K_R)},
\end{equation}
with
\begin{equation}
Z_B(K_R)\simeq4 e^{-\frac{2.5}{K_R}} K_R^{0.13}+2,
\end{equation}
where $K_R$ is the ratio of the thermal average of the $N_1$ decay
rate and Hubble parameter and is given by
\begin{equation}
K_R=\frac{M_P}{1.7\times8\pi v^2\sqrt{g_*}}
 \frac{\left( M_D^\dagger M_D\right) _{11}}{M_1}
\end{equation}
where $M_P\simeq1.22\times10^{19}$ GeV is the Planck mass. The baryon asymmetry $Y_B$ given by Eqn. (32) contain three right handed Majorana neutrino masses $M_1$, $M_2$, $M_3$ and complex phases $\xi_{ij}$. Assuming hierarchical mass spectrum of  $M_1$, $M_2$ and $M_3$ with $M_3=M_0, M_2=10^{-2} M_0, M_1=10^{-5} M_0$ and giving full variation to phases $\xi_{ij}$, we have shown in Fig. (1) $Y_B$ as a function of
the right handed Majorana neutrino mass scale, $M_0$. Fig. (1) has been plotted for
three randomly picked representative sets of the complex phases $\xi_{ij}$ or equivalently three sets of the complex orthogonal matrix $R$ to show the dependence of $Y_B$ on $M_0$ which otherwise will not be clear.  We have, also, shown the observed baryon
asymmetry as grey region between horizontal lines. It is clear from the figure that
$Y_B$ has a linear dependence on the heavy neutrino mass scale $M_0$. Also, for hybrid texture (Eqn. (9)) of the light neutrino
mass matrix, the observed baryon asymmetry predicts that the mass
scale of the right handed Majorana neutrino, $M_0$ lies in the
range $(0.3\times10^{16} \leq M_0 \leq 1.4\times10^{16})$ GeV for set $1$ and $(0.8\times10^{16} \leq M_0 \leq 3.4\times10^{16})$ GeV and $(1.8\times10^{16} \leq M_0 \leq 7.0\times10^{16})$ GeV for sets $2$ and $3$.

\section{Conclusions}
In conclusion, we examined the implications of neutrino mass
matrix with hybrid texture resulting from $Q_8$ symmetry for CP
asymmetry and Baryon Asymmetry of the Universe (BAU). We considered the Dirac neutrino mass matrix $M_D$ in the Casas-Ibarra (CI) parameterization (Eqn. (11)) alongwith the neutrino mass matrix $M_{\nu}$ with the hybrid texture obtained from $Q_8$ symmetry. With the parameterized form of the light neutrino mass matrix $M_{\nu}$ (Eqn. (12)) we find the form of the Dirac neutrino mass matrix $M_D$ corresponding to
this hybrid texture. Assuming a normal hierarchy of the right
handed Majorana neutrino masses, $M_1 \ll M_2$, $M_3$ , we
calculate the baryon asymmetry, $Y_B$, as a function of the right
handed Majorana neutrino mass scale, $M_0$. It is found that
hybrid texture of the kind considered here gives consistent value
of the baryon asymmetry, $Y_B$, for neutrino mass scale, $M_0$, of the order of $10^{16}$ GeV which is
several orders of magnitude higher than the other viable $Q_8$
scenario with two texture zeros\cite{15,16}. In particular, we have shown in Fig. (1) $Y_B$ for three different sets of complex orthogonal matrices $R$.

\newpage
\textbf{\textit{\Large{Acknowledgements}}}

 The research work of S. D. is supported
by University Grants Commission (UGC), Government of India
\textit{vide} Grant No. 34-32/2008 (SR). S. V. acknowledges the
financial support provided by University Grants Commission (UGC),
Government of India.

\newpage
\begin{figure}
\begin{center}
\epsfig{file=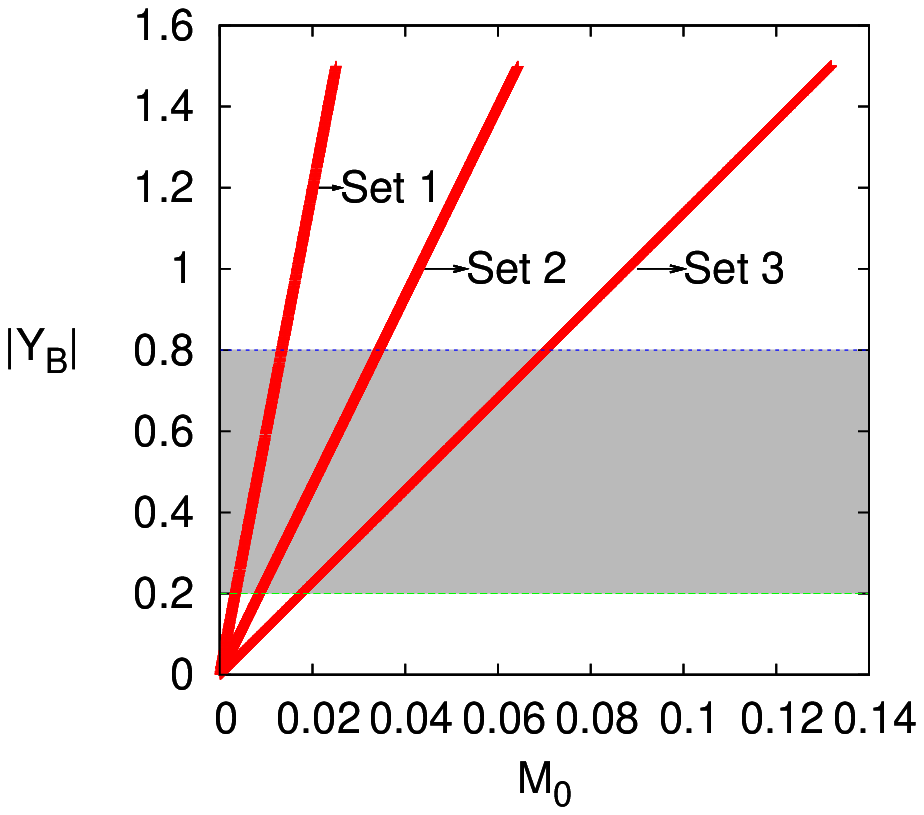, width=15.0cm, height=10.0cm}
\end{center}
\caption{Baryon Asymmetry of the Universe (BAU) $|Y_{B}|$ (in the units of $10^{-10}$), as a function of right handed Majorana neutrino mass scale $M_{0}$ (in the units of $10^{18}$ GeV). The region between the two horizontal lines is the observed $|Y_{B}|$.}
\end{figure}

\begin{thebibliography}{99}
\bibitem{1} H. C. Goh, R. N. Mohapatra and
Siew-Phang Ng, \textit{Phys. Rev.} \textbf{D 68}, 115008 (2003).
\bibitem{2}  I. de Medeiros Varzielas, S. F. King and G. G. Ross, \textit{Phys. Lett.} \textbf{B 644}, 153-157 (2007).
\bibitem{3} M. Frigerio, S. Kaneko, E. Ma and M. Tanimoto, \textit{Phys. Rev.}
\textbf{D 71}, 011901 (2005).
\bibitem{4} E. Ma and U. Sarkar, \textit{Phys. Rev. Lett.} \textbf{80}, 5716 (1998).
\bibitem{5} S. Dev, Surender Verma and Shivani Gupta, \textit{Phys. Lett.} \textbf{B 687}, 53-60 (2010). 
\bibitem{6} Paul H. Frampton, Sheldon L. Glashow and Danny
Marfatia, \textit{Phys. Lett.} \textbf{B 536}, 79 (2002).
\bibitem{7} Bipin R.
Desai, D. P. Roy and Alexander R. Vaucher, \textit{Mod. Phys.
Lett.} \textbf{A 18}, 1355 (2003).
\bibitem{8} Zhi-zhong Xing, \textit{Phys.
Lett.} \textbf{B 530} 159 (2002).
\bibitem{9} Wan-lei Guo and Zhi-zhong Xing,
\textit{Phys. Rev.} \textbf{D 67}, 053002 (2003).
\bibitem{10} Alexander Merle
and Werner Rodejohann, \textit{Phys. Rev.} \textbf{D 73}, 073012
(2006).
\bibitem{11} S. Dev,
Sanjeev Kumar, Surender Verma and Shivani Gupta,
 \textit{Nucl. Phys.} \textbf{B 784}, 103-117 (2007).
 \bibitem{12} S. Dev, Sanjeev Kumar, Surender Verma and Shivani Gupta,
 \textit{Phys. Rev.} \textbf{D 76}, 013002 (2007).
 \bibitem{13} S. Dev, Sanjeev Kumar and Surender Verma, \textit{Phys. Rev.} \textbf{D 79}, 033011 (2009), arXiv: 0901.2819 [hep-ph].
 \bibitem{14} S. Dev, Sanjeev Kumar, Surender Verma and
Shivani Gupta, \textit{Mod. Phys. Lett.} \textbf{A 24}, 2251-2261
(2009), arXiv: 0810.3080 [hep-ph].
\bibitem{15} S. Kaneko and M. Tanimoto, \textit{Phys. Lett.}
\textbf{B 551}, 127-136 (2003).
\bibitem{16} S. Kaneko, M. Katsumata and M. Tanimota,
\textit{JHEP} 0307 (2003) 025.
\bibitem{17} S. Kaneko, H. Sawanaka and M. Tanimoto, \textit{JHEP} 0508 (2005) 073.
\bibitem{18} D. Chang, W. Y. Keung and G. Senjanovic, \textit{Phys. Rev.} \textbf{D 42}, 1599 (1990).
\bibitem{19} D. Chang, W. Y. Keung, S. Lipovaca and G. Senjanovic, \textit{Phys. Rev. Lett.} \textbf{67}, 953 (1991).
\bibitem{20} P. H. Frampton and T. W. Kephart, \textit{Int. J. Mod. Phys.} \textbf{A 10}, 4689 (1995).
\bibitem{21} P. H. Frampton and O. C. W. Kong, \textit{Phys. Rev. Lett.} \textbf{75}, 781 (1995).
\bibitem{22} P. H. Frampton and A. Rasin, \textit{Phys. Lett.} \textbf{B 478}, 424 (2000). 
\bibitem{23} P. Minkowski, \textit{Phys. Lett.} \textbf{B 67}, 421 (1977).
\bibitem{24} T. Yanagida, in \textit{Proceedings of the Workshop on the Unified
Theory and the Baryon Number in the Universe} (O. Sawada and A.
Sugamoto, eds.), KEK, Tsukuba, Japan, 1979, p. 95.
\bibitem{25} M. Gell-Mann,
P. Ramond, and R. Slansky, \textit{Complex spinors and unified
theories}, in \textit{Supergravity} (P. van Nieuwenhuizen and D.
Z. Freedman, eds.), North Holland, Amsterdam, 1979, p. 315.
\bibitem{26} S. L. Glashow, \textit{The future of elementary particle physics}, in
\textit{Proceedings of the 1979 Cargese Summer Institute on Quarks
and Leptons} (M. Levy, J. L. Basdevant, D. Speiser, J. Weyers, R.
Gastmans and M. Jacob, eds.), Plenum Press, Newyork, 1980, pp.
687-713.
\bibitem{27} Werner Rodejohann, \textit{Pramana} \textbf{72}, 217-227 (2009).
\bibitem{28} J. A. Casas and A. Ibarra, \textit{Nucl. Phys.} \textbf{B 618}, 171 (2001).
\bibitem{29} M. C. Gonzalez-Garcia and Michele Maltoni, \textit{Phys.
Rept.} \textbf{460} (2008) 1-129, arXiv: 0704.1800v2 [hep-ph].
\bibitem{30} V. A. Kuzmin, V. A. Rubakov and M. E. Shaposhnikov, \textit{Phys. Lett.}
\textbf{B 155}, 36 (1985).
\bibitem{31} M. Kobayashi and T. Maskawa, \textit{Prog. Theor.
Phys.} \textbf{49}, 652-657 (1973).
\bibitem{32} V. A. Rubakov and M. E. Shaposhnikov, \textit{Usp. Fiz. Nauk} \textbf{166},
493-537 (1996).
\bibitem{33} Mark Trodden, \textit{Rev. Mod.
Phys.} \textbf{71}, 1463-1500 (1999).
\bibitem{34} Ian Affleck and Michael Dine, \textit{Nucl. Phys.}
\textbf{B 249}, 361 (1985).
\bibitem{35} Michael Dine, Lisa Randall and Scott D. Thomas, \textit{Nucl. Phys.}
\textbf{B 458}, 291-326 (1996).

\bibitem{36} A. Yu. Ignatiev, N. V. kasnikov, V. A. Kuzmin and A. N. Tavkhelidze,
\textit{Phys. Lett.} \textbf{B 76}, 436-438 (1978).
\bibitem{37} Motohiko
Yoshimura, \textit{Phys. Rev. Lett.} \textbf{41}, 281-284 (1978).

\bibitem{38} D. Toussaint, S. B. Treiman, Frank Wilczek and A. Zee, \textit{Phys. Rev.} \textbf{D 19},
1036-1045 (1979).
\bibitem{39} Savas Dimopoulos and Leonard Susskind,
\textit{Phys. Rev.} \textbf{D 18}, 4500-4509 (1978).
\bibitem{40} John R.
Ellis, Mary K. Gaillard and Dimitri V. Nanopoulos, \textit{Phys.
Lett.} \textbf{B 80}, 360 (1979).
\bibitem{41} Steven Weinberg, \textit{Phys.
Rev. Lett.} \textbf{42}, 850-853 (1979).

\bibitem{42} Motohiko Yoshimura, \textit{Phys. Lett.} \textbf{B 88}, 294 (1979).
\bibitem{43} Stephen M. Barr, Gino Segre and
H. Arthur Weldon, \textit{Phys. Rev.} \textbf{D 20}, 2494 (1979).
\bibitem{44} Dimitri V. Nanopoulos and Steven Weinberg, \textit{Phys. Rev.}
\textbf{D 20}, 2484 (1979).
\bibitem{45} Asim Yildiz and Paul Cox,
\textit{Phys. Rev.} \textbf{D 21}, 906 (1980).

\bibitem{46} M. Fukugita and T. Yanagida, \textit{Phys. Lett.} \textbf{B
174}, 45 (1986).
\bibitem{47} S. Yu. Khlebnikov and M. E. Shaposhnikov, \textit{Nucl. Phys.} \textbf{B
308}, 885-912 (1988).

\bibitem{48} S. Davidson, E. Nardi, Y. Nir, \textit{Phys. Rept.} \textbf{466}, 105-177 (2008); arXiv: 0802.2962v3 [hep-ph].

\bibitem{49} Srubabati Goswami, Subrata Khan, Werner Rodejohann, \textit{Phys. Lett.} \textbf{B 680}, 255-262 (2009); arXiv: 0905.2739v2 [hep-ph].
\bibitem{50} J. A. Harvey and M. S. Turner, \textit{Phys. Rev.} \textbf{D
42}, 3344 (1990).
\bibitem{51} S. Y. Khlebnikov and S. E. Shaposhnikov,
\textit{Nucl. Phys.} \textbf{B 308}, 169 (1998).
\bibitem{52} M. Luty, \textit{Phys. Rev.} \textbf{D 45}, 455 (1992).
\bibitem{53} M. Plumacher, \textit{Z. Phys.} \textbf{C 74} 549 (1997).
\bibitem{54} Evgeny Kh. Akhmedov and Werner Rodejohann, \textit{JHEP} \textbf{0806}, 106 (2008).
\bibitem{55} W. Buchmuller, P. Di Bari, M. Plumacher, \textit{Annals Phys.} \textbf{315}, 305-351 (2005).
\bibitem{56} Steve Blanchet and Pasquale Di Bari, \textit{JCAP} \textbf{0703}, 018 (2007).

\end{thebibliography}
\end{document}